\documentclass[%
 reprint,
superscriptaddress,
 amsmath,amssymb,
 aps,
]{revtex4-2}

\usepackage{graphicx}
\usepackage{dcolumn}
\usepackage{bm}
\usepackage[dvipsnames]{xcolor}

\begin{document}

\preprint{APS/123-QED}

\title{Challenges and mitigation pathways in coating silver nanowire networks\\ with metallic oxides by RF magnetron sputtering}

\author{Amaury Baret}
\affiliation{%
Department of Physics, SPIN, University of Liège,\\
Allée du Six Août 19, B–4000 Liège, Belgium\relax
}%
 \email{abaret@uliege.be}

\author{Ambreen Khan}%
\affiliation{%
Univ. Bordeaux, CNRS, Bordeaux INP, ICMCB, UMR 5026, 87 Av. du Dr. Schweitzer, F-33600 Pessac, France
}%
\affiliation{%
Univ. Grenoble Alpes, CNRS, Grenoble INP, LMGP, F-33800 Grenoble, France
}%

\author{Sude Akin}
\affiliation{%
Univ. Bordeaux, CNRS, Bordeaux INP, ICMCB, UMR 5026, 87 Av. du Dr. Schweitzer, F-33600 Pessac, France
}%

\author{Lionel Teulé-Gay}
\affiliation{%
Univ. Bordeaux, CNRS, Bordeaux INP, ICMCB, UMR 5026, 87 Av. du Dr. Schweitzer, F-33600 Pessac, France
}%
\author{Daniel Bellet}
\affiliation{%
Univ. Grenoble Alpes, CNRS, Grenoble INP, LMGP, F-33800 Grenoble, France
}%

\author{Aline Rougier}
\affiliation{%
Univ. Bordeaux, CNRS, Bordeaux INP, ICMCB, UMR 5026, 87 Av. du Dr. Schweitzer, F-33600 Pessac, France
}%

\author{Ngoc Duy Nguyen}
\affiliation{%
Department of Physics, SPIN, University of Liège,
Allée du Six Août 19, B–4000 Liège, Belgium
}

\date{\today}

\begin{abstract}
As silver nanowire (AgNW) networks reach increasing technological maturity, research efforts are progressively shifting toward their integration into functional devices. In this context, it is essential to assess how thin film coating processes affect the structural and functional integrity of these transparent conducting networks. Radio Frequency (RF) magnetron sputtering is among the most widely used and industrially scalable deposition techniques, making a detailed understanding of its impact on AgNW networks particularly critical.
In this work, we experimentally investigate the degradation of AgNW networks observed under specific RF magnetron sputtering regimes. By varying deposition time, oxygen partial pressure, target material, buffer layers and plasma power, we analyze how sputtering conditions influence the electrical, morphological, and structural properties of the networks. Based on these observations, we identify viable strategies to mitigate or suppress network degradation, thereby enabling safer and more reliable coating protocols. These results provide practical guidelines for the integration of AgNW networks into multilayer device architectures.

\end{abstract}

\maketitle

\section{Introduction}
Metallic nanowire networks have attracted considerable attention owing to their exceptional electro-optical performance, mechanical flexibility, and compatibility with scalable, low-cost fabrication methods such as roll-to-roll coating~\cite{nguyenAdvances2022e}. Among these, silver nanowire (AgNW) networks stand out as particularly promising due to their mature synthesis routes, superior chemical stability compared to copper, and greater availability than nickel~\cite{mauryaRecent2025}.

The optical and electrical properties of AgNW networks, as well as their dependence on structural parameters and fabrication methods, are now relatively well understood~\cite{huoOneDimensional2019a,belletTransparent2017d}. Consequently, recent research has increasingly focused on integrating AgNWs into multilayer or hybrid devices, where they must withstand additional processing steps~\cite{liRecent2025a,khanSilver2024,baretLowemissivity2025b,chengOnedimensional2022,felipegerleinPhotonic2024}. In such contexts, understanding how subsequent deposition processes affect network integrity is crucial. Among the various industrially relevant coating techniques, Radio Frequency (RF) magnetron sputtering is one of the most widespread technological methods used to deposit functional layers such as oxides, metals, and nitrides~\cite{borowskiRecent2025,gudmundssonPhysics2020,kellyMagnetron2000}.

Comprehensive treatments of the physical principles and technological implementation of RF magnetron sputtering are available in the dedicated literature~\cite{hipplerLow2008,liebermanPrinciples2003}. Here, we provide a concise, high-level description to introduce the concepts necessary for the following discussion. RF magnetron sputtering is a physical vapor deposition technique in which a plasma is used to erode a target material. The sputtered species subsequently propagate through the vacuum chamber, and a fraction of them reaches the substrate, where film growth occurs.
Owing to its compatibility with both metallic and dielectric targets, arc-free process, low substrate temperature, scalability to large-area deposition, and relatively low operational cost, RF magnetron sputtering is one of the most widely employed techniques for thin film fabrication. Nevertheless, it is well established that this process can induce substrate damage, particularly during the deposition of insulating oxide films, where energetic ion bombardment plays a critical role~\cite{aydinSputtered2021,yangOrigin2024}.

Substrate degradation during reactive sputtering is generally attributed to physical damage from energetic particles capable of inducing sputtering or resputtering from the substrate surface~\cite{mahieuModeling2009,carciaInfluence2003}.
A variety of energetic species may contribute to this etching process, including ions originating from the plasma, positive ions emitted or backscattered from the target, reflected neutral atoms, and negative ions formed at the target surface \cite{aydinSputtered2021}. Among these, negative oxygen ions generated at the target surface are particularly critical, as they can be accelerated across the cathode sheath and acquire kinetic energies of several hundred electronvolts. As a result, they are widely considered the most damaging species for substrate integrity under reactive sputtering conditions~\cite{ellmerReactive2012}. Their kinetic energy is primarily determined by the sheath voltage, whereas their flux depends on both discharge conditions and target-related properties~\cite{mahieuModeling2009}. Experimental studies have shown that O$^{-}$ ions constitute the most abundant \emph{high-energy} negative ion species in reactive sputtering plasmas, and that their impact on the growing film can lead to resputtering, surface erosion, and morphological degradation \cite{kesterMicroeffects1993}.

The severity of this effect is strongly material-dependent. Although the detailed mechanisms governing the generation of high-energy negative oxygen ions remain an active area of research \cite{deplaMeasurement2024}, it is well established that negative-ion emission is closely linked to the chemical state of the target surface \cite{welzelNegative2012}. In particular, the presence of an oxidized surface layer is a prerequisite for significant O$^{-}$ emission: multiple studies report substantially higher negative oxygen ion yields from oxidized targets compared to metallic ones, while negligible emission is observed in the absence of surface oxygen \cite{deplaMeasurement2024a,tucekSecondaryelectron1996}. 

Consequently, if negative oxygen ions are indeed the dominant cause of the observed degradation, metallic targets operated under oxygen-free conditions are not expected to generate substantial negative oxygen ion fluxes and should therefore exhibit limited substrate degradation from this mechanism. When oxygen is introduced into the discharge, however, target oxidation becomes possible, and material-dependent differences in negative oxygen ion yield emerge. These differences are expected to correlate with the thermodynamic stability of the target oxide, since the oxide formation enthalpy governs the ease and stability of surface oxidation under reactive conditions. Materials forming more stable oxides are therefore more likely to sustain oxidized surface states and, in turn, higher O$^{-}$ emission. While this qualitative trend is supported by experimental observations, a direct and quantitative relationship between oxide formation enthalpy and negative oxygen ion yield has not yet been firmly established~\cite{deplaMagnetron2009}.

In the specific case of AgNW networks, the effects of sputtering remain poorly understood and seldom addressed. A few studies report successful oxide coatings with materials such as ZnO~\cite{singhSilver2016a,sreedharFacile2019,kimHighly2013a}, silver~\cite{leeEnhancing2024} and AZO~\cite{wuAZO2019}, while Khan et al. observed degradation of AgNWs during the deposition of V$_2$O$_5$ films, although the underlying mechanisms of this degradation were not clearly identified~\cite{khanSilver2024}. Yet, understanding this instability is essential for the reliable integration of AgNW electrodes into oxide-based devices such as solar cells or chromogenic devices.  

Past research on the RF sputtering damage of oxides has shown that several approaches can mitigate ion-induced degradation during sputtering~\cite{reddyHolistic2022,yuanMechanistic2024,suemoriAssessment2023}:  
\begin{itemize}
    \item Lowering target voltage to reduce O$^-$ acceleration 
    \item Introducing buffer layers that screen the deleterious effects of the incidents ions.  
    \item Increasing chamber pressure to enhance scattering of O$^-$ ions, thereby lowering their energy at the substrate, at the expense of slower deposition rates~\cite{anderssonEnergy2006}.  
    \item Employing off-axis sputtering geometries, which prevent direct ion bombardment of the substrate.
    \item Varying the target nature (metallic or oxide) to reduce the creation rate of negative oxygen ions.  
    
\end{itemize}  

The sparse experimental details in the existing literature on AgNW sputtering suggest that some of these mitigation strategies may already have been implemented implicitly. For example, off-axis sputtering could fully suppress ion bombardment, enabling the successful deposition of oxide films, although to the cost of a drastic reduction of the deposition rate and increased waste of raw materials. Systematic evaluation of these strategies would accelerate the development of optimized processes for transparent conductive oxide coatings on AgNW networks, a critical step toward the deployment of this hybrid material in advanced optoelectronic devices~\cite{sekkatEnhanced2024b}.  
Khan et al.~\cite{khanSilver2024} have shown that a buffer layer consisting of ALD-deposited ZnO or SnO$_2$ can protect the AgNW network during the sputtering, provided it is sufficiently thick (thickness values larger than 25 nm were reported). 

In this work, we investigate the impact of RF magnetron sputtering on the stability of AgNW networks. We first examine the degradation kinetics through in situ electrical resistance monitoring, providing time-resolved insight into the failure process. We then analyze the respective roles of target material (metallic vs. oxide) and reactive plasma conditions in order to identify the dominant degradation pathways, and correlate these effects with the electrical, structural, and morphological evolution of the networks. Finally, we assess the effectiveness of a ZnO buffer layer as a stabilization strategy during sputtering. The influence of argon pressure on network stability is additionally discussed in the Supporting Information (see Fig.~\ref{fig:ArPressureXRDSEM}).

\section{Experimental methods}
\subsection{AgNW networks deposition}
Corning glass substrates, each measuring 25 mm $\times$ 25 mm, were cleaned prior to silver nanowire deposition. This includes initial cleaning with acetone, followed by ultrasonic treatment in ethanol and then in deionized water for 15 min, and finally drying using a nitrogen gas stream. Silver nanowires sourced from Protavic (70 nm in diameter and 10 µm in length) were diluted with isopropanol to reach a final concentration of 0.23 g/L. The deposition of AgNW onto the glass substrates was carried out via spray coating with an airbrush system, connected to a programmable ‘PAC Display Runtime Professional’ software. The solution was inserted into the reservoir with a syringe, and then sprayed using a carrier gas under pressure, which allows adjustment of the flow rate. During the deposition, an airbrush moves in two directions, creating a uniform, repeating pattern across the target surface. The substrates were placed on an aluminum plate heated at 110°C, which facilitates the evaporation of solvent after each deposition cycle, leaving only the nanowires on the substrates. The targeted resistance for the sample was controlled by in situ measurement of resistance using a home-made monitor (glass with two parallel contacts using silver paste on opposite sides) connected to a multimeter. The desired resistance of the AgNW network was obtained by monitoring the resistance during spray deposition. The samples were thermally annealed for 1 h at 200°C to optimize the junction between AgNWs and thus improve conductivity.

\subsection{SALD coating}
The protective oxide coatings of zinc oxide (ZnO) or tin(IV) oxide (SnO$_2$) on AgNW networks were deposited by atmospheric-pressure spatial atomic layer deposition (AP-SALD). The SALD setup uses a specially designed proximity-type deposition head. The head features multiple parallel gas channels for each precursor, separated by inert nitrogen gas. The nitrogen streams effectively isolate the reactant vapors and eliminate any risk of gas-phase mixing or unwanted reactions before reaching the substrate surface. The precursors used for ZnO coating were diethyl zinc and water, whereas tin(II) acetylacetonate and water were employed as precursors for SnO$_2$. The deposition was carried out at a substrate temperature of 200 °C and ambient atmospheric pressure. The deposition head moves linearly back and forth over the stationary substrate, maintaining a 120 $\mu$m head-to-substrate gap. Each cycle follows these steps: the metal precursor was first supplied to saturate the surface, followed by purging of N$_2$ to remove by-products; water vapor was then introduced as the oxidant to complete the reaction; and a final N$_2$ purge removed residual water and reaction products. The resulting growth per cycle translated to effective deposition rates of approximately 0.26 nm/s for ZnO and 0.02 nm/s for SnO$_2$. By varying the total number of cycles, the oxide coating thickness encapsulating the silver nanowires was controlled at 30 nm, verified by spectroscopic ellipsometry on a reference bare glass.

\subsection{RF magnetron sputtering and in situ resistance measurements}

Thin films of WO$_3$ and NiO were deposited by reactive RF magnetron sputtering using a MP700S system (Plassys). WO$_3$ films were obtained from either metallic W or ceramic WO$_3$ targets, while NiO films were deposited from a metallic Ni target. Copper and stainless steel (inox) depositions were carried out in a NEXDEP-class sputtering system (Angstrom Engineering). All targets were 2~in. in diameter. 
Prior to each deposition, the chamber base pressure was reduced below $3\times10^{-5}$~Pa. Argon was used as the sputtering gas, and oxygen was introduced as the reactive gas when required. The total pressure, oxygen partial pressure, and RF power were adjusted depending on the experiment and are specified in the corresponding figure captions and discussion sections. All depositions were performed in an on-axis configuration without substrate rotation.
For in situ electrical monitoring, AgNW samples were electrically contacted inside the sputtering chamber using vacuum feedthroughs. The two ends of each sample were connected with conductive tape to copper leads wired to an external source meter (Keithley 2450).
The substrate temperature during sputtering was monitored using thermocouples positioned in close proximity to the sample holder.

\subsection{Structural, morphological, and electrical characterization}

Surface morphology was examined by field-emission scanning electron microscopy (FEG-SEM) using a Gemini 300 microscope (Zeiss) and a Pioneer Two system (Raith).
Grazing-incidence X-ray diffraction measurements were performed using Bruker D8 diffractometers (Advance and Discover Twin-Twin systems) with Cu K$\alpha$ radiation ($\lambda = 0.1541$~nm).

\section{Results and discussion}

\subsection{In situ resistance evolution and failure kinetics}
Understanding the kinetics of AgNW degradation during RF sputtering is essential for disentangling the competing processes of deposition and etching. In principle, one might expect that there exists an optimal exposure time at which the coating layer begins to form while the etching of the nanowires remains minimal. Identifying such a characteristic timescale is also critical for designing the remainder of the experimental plan, since the deposition time must be first determined when exploring the influence of other sputtering parameters.

\begin{figure*}
    \centering
    \includegraphics[width=0.85\linewidth]{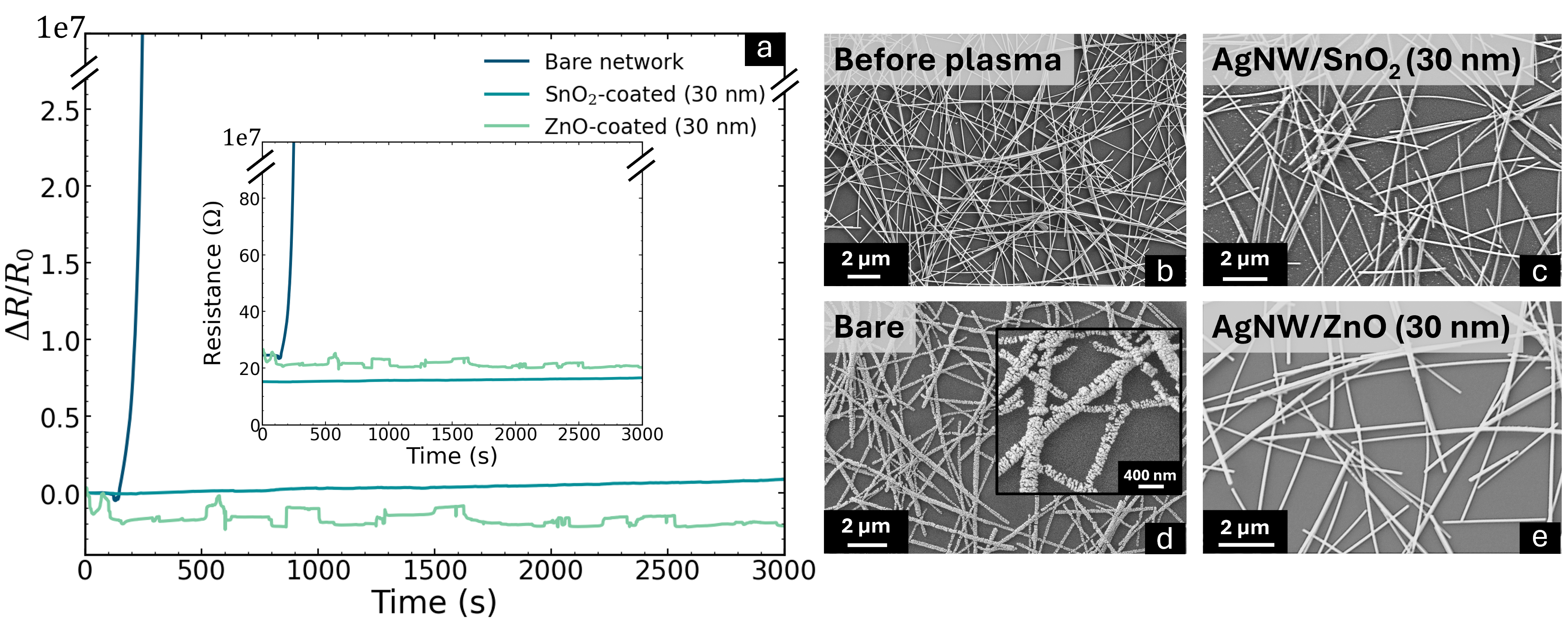}
    \caption{\textit{(a)} Relative electrical resistance variation of AgNW networks on glass as a function of sputtering time for bare and coated samples. RF magnetron sputtering was performed at a plasma power of 200~W, an oxygen partial pressure of 10\%, and a total argon pressure of 2~Pa. The inset shows the corresponding absolute resistance evolution. 
\textit{(b-e)} SEM images of AgNW networks: \textit{(b)} before RF sputtering and \textit{(c-e)} after sputtering. The samples shown in \textit{(c)} and \textit{(e)} were protected by a 30~nm oxide layer deposited by SALD, consisting of SnO$_2$ and ZnO, respectively. \textit{(d)} Bare AgNW network after sputtering.
}
    \label{fig:cinetiqueRSEM}
\end{figure*}

To probe the time evolution of the network's integrity, we monitored the in situ electrical resistance of AgNW samples during RF sputtering. A power of 200 W with a partial oxygen pressure of 10\% and a total pressure of 2 Pa was used. Figure \ref{fig:cinetiqueRSEM}a shows the resistance evolution for both bare and oxide-coated networks. ZnO and SnO$_2$ buffer layers are included here as representative oxide barriers to probe the origin of sputtering-induced degradation. The results reveal that any degradation, when present, occurs extremely rapidly: within the first few minutes of exposure. This timescale is incompatible with typical deposition durations, which extend up to several hours. Consequently, tuning the sputtering time cannot mitigate degradation once the process has started.

These measurements, shown in Fig.~\ref{fig:cinetiqueRSEM}a, already suggest that suitable oxide buffer layers can suppress or eliminate the degradation pathway. SEM imaging supports this conclusion (see Figure~\ref{fig:cinetiqueRSEM}b): bare AgNW networks exhibit severe structural damage after sputtering, with nanowires fractured into isolated segments. In contrast, samples coated with 30 nm-thick SnO$_2$ or ZnO protective coatings show no detectable morphological deterioration under identical sputtering conditions, and their electrical resistance remains stable (or even improves marginally). Both ZnO and SnO$_2$ exhibit similar qualitative behavior as coating materials. Consequently, ZnO will serve as the model protective layer (see Section \ref{sec:coatings}).

It is noteworthy that the temperature increase measured at the stage during the deposition process never exceeded 80°C, indicating that the cause of the observed degradation is not thermal given that AgNWs of these dimensions are thermally stable up to 250°C~\cite{baltyInsight2024a}.

In summary, in situ resistance monitoring provides direct insight into the kinetics of AgNW degradation and shows that the characteristic timescale of damage is much shorter than typical sputtering durations, making temporal adjustment an ineffective mitigation strategy.

\subsection{Coupled influence of oxygen partial pressure and target material on AgNW stability}

To elucidate the mechanisms responsible for AgNW degradation during RF magnetron sputtering, we investigated the combined influence of oxygen partial pressure and target material. Based on the literature and the observations reported above, we hypothesize that the stability of AgNW networks is governed by the generation of energetic oxygen negative ions at the target surface. The efficiency of this process is expected to depend jointly on the oxygen availability in the plasma and on the surface chemistry of the sputtering target.

We first establish the role of oxygen by examining sputtering conditions in its absence. Bare AgNW networks were exposed to copper and stainless steel (inox) targets under identical plasma conditions (2~Pa Ar, 200~W RF power). These two materials were selected because they are known to produce markedly different sputtering environments: copper exhibits a relatively high sputtering yield~\cite{Kurt}, whereas stainless steel is significantly more resistant to sputtering~\cite{wayner.Ion1976}.

\begin{figure*} 
\centering 
\includegraphics[width=0.85\linewidth]{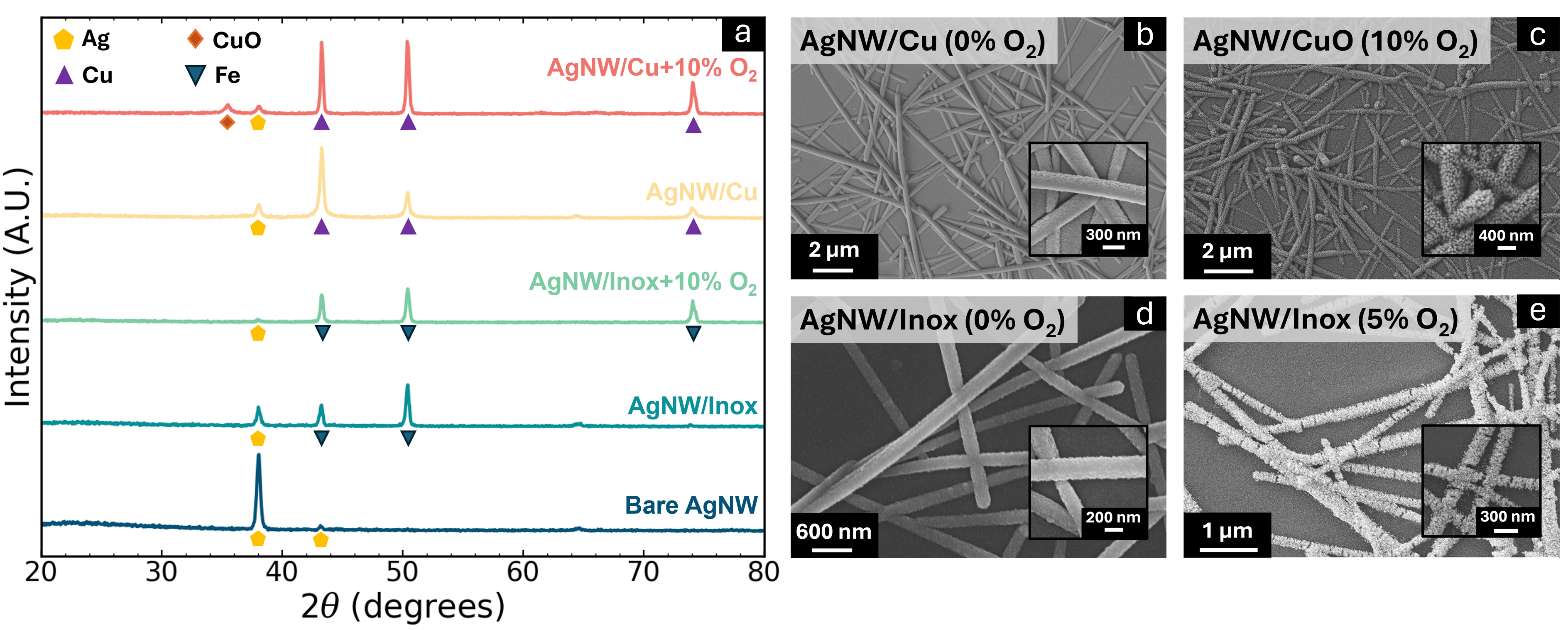} 

\caption{\textit{(a)} Grazing incidence XRD patterns of AgNW networks on glass before sputtering (Bare) and after RF magnetron sputtering using different target materials and oxygen partial pressures. Crystalline peak positions are indicated by symbols. \textit{(b-e)} SEM images of AgNW networks after sputtering under the corresponding conditions: \textit{(b)} Cu target without O$_2$ in the chamber, \textit{(c)} Cu target with 10\% O$_2$, \textit{(d)} stainless steel (inox) target without O$_2$, and \textit{(e)} inox target with 5\% O$_2$. }
\label{fig:O2XRDSEM} 

\end{figure*}

As shown in Fig~\ref{fig:O2XRDSEM}, under oxygen-free conditions, both targets yield minimal structural modification of the AgNWs. XRD patterns remain essentially unchanged, SEM images reveal conformal coatings without visible damage, and the electrical resistance remains low (Table~\ref{tab:summary_targets}). These observations demonstrate that plasma exposure alone, with the plasma confined near the target, is insufficient to induce degradation of the nanowire network.

A notably different behavior emerges upon introducing oxygen into the discharge. When sputtering with the inox target at 5-10\% O$_2$, catastrophic degradation is observed: the network resistance becomes infinite, SEM images show advanced fragmentation of the nanowires into isolated segments, and the Ag diffraction peaks strongly diminish or disappear entirely. The absence of detectable Ag$_2$O peaks indicates that the dominant degradation pathway is not simple chemical oxidation of silver but is instead more consistent with energetic ion-induced etching, most plausibly driven by negatively charged oxygen species generated at the target.

In contrast, sputtering with the copper target under otherwise comparable oxygen-containing conditions preserves the structural and electrical integrity of the AgNW network. Although CuO peaks appear in XRD when oxygen is present, SEM observations show no nanowire fragmentation, and the electrical resistance remains finite and even decreases after deposition. Because undoped CuO is highly resistive, this reduction cannot originate from current shunting through the coating and instead confirms that the AgNW percolation network remains intact.

Taken together, these results demonstrate that oxygen presence alone does not guarantee degradation. Rather, the severity of damage depends critically on the target material, supporting the hypothesis that the generation efficiency of energetic oxygen negative ions is strongly target-dependent.

\begin{figure*} 
\centering 
\includegraphics[width=0.85\linewidth]{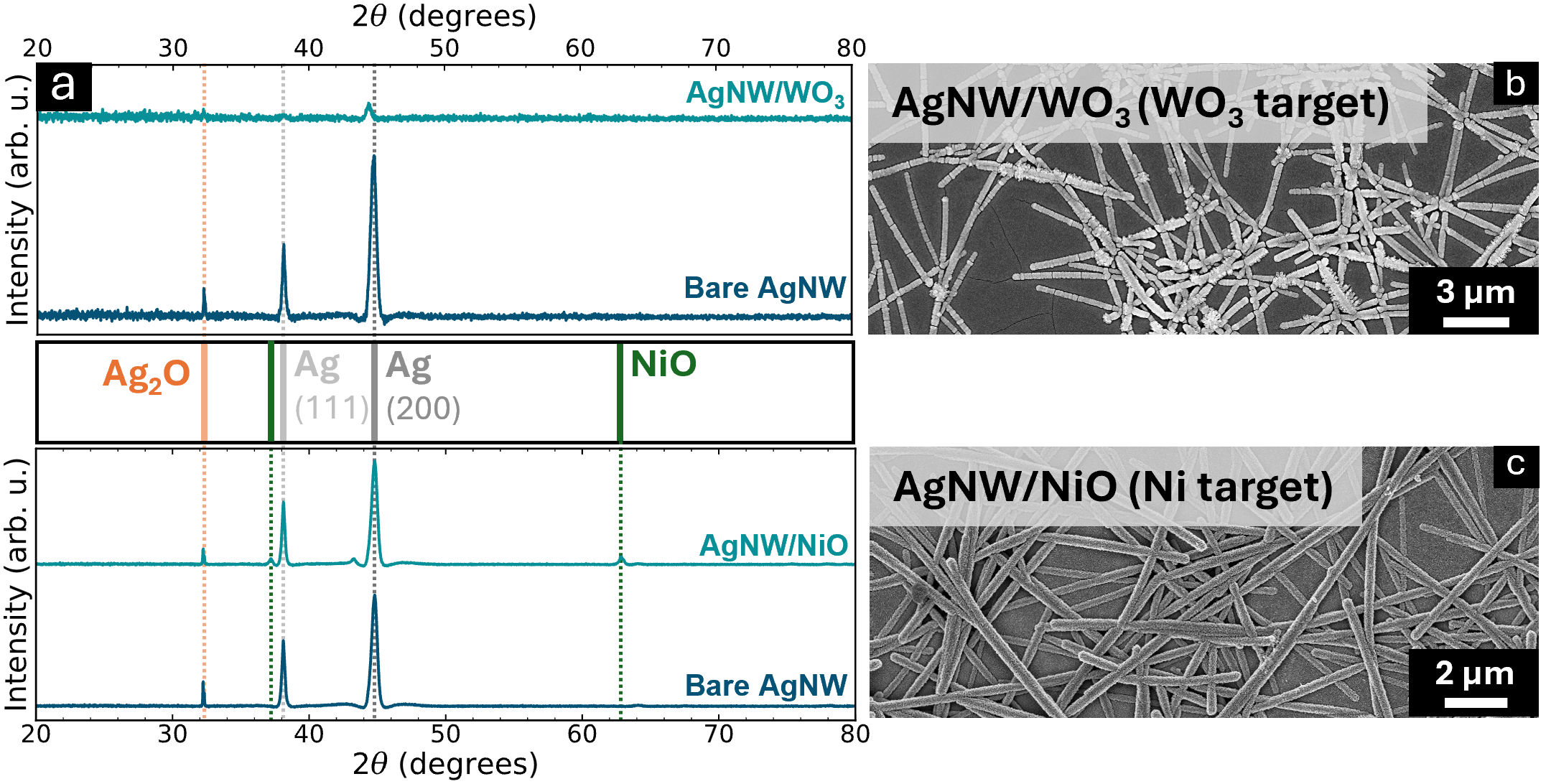} 
\caption{\textit{(a)} Grazing incidence XRD patterns of bare AgNW networks on glass before (dark blue) and after (turquoise) RF magnetron sputtering using different targets. The top panel corresponds to sputtering with a WO$_3$ target at 2\% oxygen partial pressure, 4~Pa total pressure, and 75~W RF power, while the bottom panel corresponds to sputtering with a Ni target at 1\% oxygen partial pressure, 4~Pa total pressure, and 100~W RF power. \textit{(b,c)} SEM images of AgNW networks after sputtering using \textit{(b)} a WO$_3$ target and \textit{(c)} a Ni target, respectively. }
\label{fig:TargetXRDSEM} 
\end{figure*}

To further probe the role of target chemistry, AgNW networks were sputter-coated using a WO$_3$ (oxide) target and a metallic Ni target under reactive conditions optimized for functional oxide growth (2\% O$_2$ for WO$_3$, 1\% O$_2$ for NiO). The total pressure was fixed at 4~Pa, with RF powers of 75~W and 100~W, respectively, chosen to reproduce realistic deposition conditions for device integration~\cite{khanSilver2024}.

As displayed in Fig.~\ref{fig:TargetXRDSEM}, the WO$_3$ target induces severe degradation despite the relatively low oxygen fraction. XRD measurements show an almost complete disappearance of the Ag peak, while SEM imaging reveals extensive cracking and fragmentation of the nanowires. The network becomes electrically insulating, indicating a loss of percolation. This result demonstrates that even modest oxygen partial pressures can trigger crucial damage when the target efficiently generates oxygen negative ions. The deposited WO$_3$ layer is amorphous under these conditions and therefore does not produce distinct diffraction peaks.
In sharp contrast, sputtering with the metallic Ni target preserves the AgNW network. The Ag diffraction peaks remain largely unchanged, crystalline NiO peaks are present, and SEM images show a conformal coating without visible nanowire damage. The electrical resistance remains finite, confirming that electrical percolation is maintained.

The combined observations reveal a clear trend: targets that either are oxides or readily form stable surface oxides under reactive conditions are systematically associated with severe AgNW degradation. Conversely, metallic targets with lower effective oxygen affinity under the investigated conditions (Cu and Ni) do not induce observable damage.

Importantly, this framework also explains the behavior of metallic tungsten observed in the Ar total pressure experiments reported in the supplementary materials: although tungsten is introduced as a metallic target, its strong thermodynamic affinity for oxygen promotes rapid formation of an oxidized surface layer during reactive sputtering. As a result, the target effectively behaves as a partially oxidized surface capable of efficiently generating energetic oxygen negative ions, thereby leading to nanowire degradation. This observation highlights that the operative parameter is not simply whether the target is nominally metallic or oxidic, but rather the extent of surface oxidation under plasma conditions.

In all experimental configurations, the substrate temperature remained below 80~$^\circ$C. Such temperatures are insufficient to induce thermally driven morphological degradation of silver nanowires, further supporting the conclusion that the observed damage originates from sputtering-related energetic species rather than from thermal effects.
Collectively, the results establish that AgNW degradation during RF magnetron sputtering is governed by a coupled oxygen–target effect. Oxygen in the discharge is a necessary but not sufficient condition for damage; the decisive factor is the ability of the target surface to generate high-energy oxygen negative ions. Targets that are oxidic or that easily form stable surface oxides promote high negative-ion fluxes and systematically lead to complete nanowire fragmentation. In contrast, metallic targets that remain weakly oxidized under the investigated conditions produce significantly less damage. This is summarized in Table~\ref{tab:summary_targets}, where structural (XRD) and electrical measurements provide consistent and complementary evidence for the proposed degradation mechanism.
This mechanistic understanding provides practical guidelines for the integration of AgNW electrodes in sputtered multilayer stacks. In particular, careful selection of target materials and control of reactive conditions are essential, and additional mitigation strategies, such as protective buffer layers, remain necessary when oxide sputtering cannot be avoided.

\begin{table*}[htbp]
\centering
\caption{Summary of AgNW network stability under RF magnetron sputtering with different targets and oxygen partial pressures. The oxide formation enthalpy $\Delta H_f$ is provided when relevant to assess the tendency of the target to oxidize and generate negative oxygen ions. High magnitude, negative $\Delta H_f$ values correspond to the most stable oxide under sputtering conditions.}
\label{tab:summary_targets}
\begin{tabular}{l l c c c c}
\hline
Sample & Target & $\Delta H_f$ (kJ/mol) & $p_{\mathrm{O}_2}$ (\%) & Ag XRD peak & $\Delta R / R_0$ \\
\hline
Bare AgNWs & Inox & N/A & 0  & Present        & $-0.23$ \\
Bare AgNWs & Inox & N/A & 5  & Absent         & $\infty$ \\
Bare AgNWs & Inox & N/A & 10 & Absent         & $\infty$ \\
Bare AgNWs/ZnO (30 nm) & Inox & N/A & 10 & Present         & -0.2 \\

\hline
Bare AgNWs & Cu   & $-156.06$ (Cu$_2$O) & 0  & Present        & $-1.14$ \\
Bare AgNWs & Cu   & $-156.06$ (Cu$_2$O) & 10 & Present        & $-0.06$ \\
\hline
Bare AgNWs & Ni   & $-239.7$ (NiO)      & 1  & Present        & 3.1 \\
AgNWs/ZnO (30 nm) & Ni & $-239.7$ (NiO)  & 1  & Present  & 1.8 \\
\hline
Bare AgNWs & W    & $-842.9$ (WO$_3$)   & 12  & Almost absent  & $\infty$ \\
\hline
Bare AgNWs & WO$_3$ & N/A               & 2  & Almost absent  & $\infty$ \\
AgNWs/ZnO (30 nm) & WO$_3$ & N/A               & 2  & Present  & 10$^3$ \\
\hline
\end{tabular}
\end{table*}

\subsection{Protective oxide buffer layers for mitigating sputtering-induced damage in AgNW networks}
\label{sec:coatings}
To assess the protective capability of oxide buffer layers under conditions representative of functional device fabrication, AgNW networks were sputter-coated using either a metallic Ni target or a WO$_3$ target, with and without an intermediate ZnO buffer layer. Distinctly different behaviors are observed depending on the target chemistry.

\begin{figure*}
    \centering
    \includegraphics[width=0.85\linewidth]{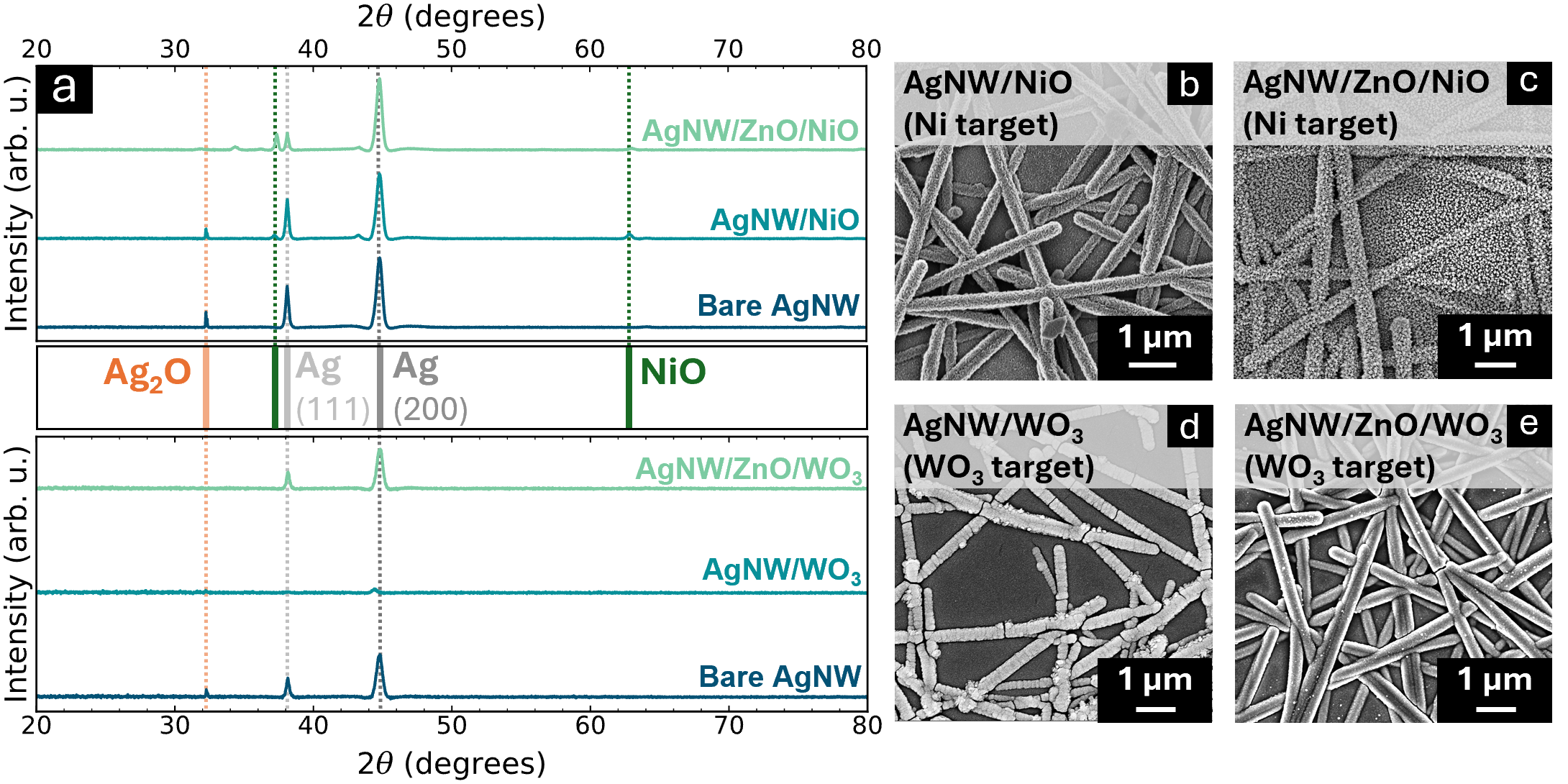}
    \caption{Effect of oxide buffer layers on the structural stability of AgNW networks during RF magnetron sputtering. (a) XRD patterns of AgNW networks after coating, with NiO (top) and WO$_3$ (bottom) sputtering, with and without ZnO (30 nm) buffer layer. Characteristic Ag diffraction peaks remain clearly visible after NiO deposition, whereas they disappear after direct WO$_3$ sputtering, indicating severe network degradation. (b–e) SEM micrographs of AgNW networks after RF sputtering of oxide overlayers: (b) NiO deposited directly on bare AgNWs, (c) NiO deposited on ZnO-buffered AgNWs, (d) WO$_3$ deposited directly on bare AgNWs, and (e) WO$_3$ deposited on ZnO-buffered AgNWs. Direct WO$_3$ deposition leads to pronounced nanowire fragmentation, while the introduction of a thin ZnO buffer layer effectively preserves network morphology during subsequent oxide growth.}
    \label{fig:coatings}
\end{figure*}

As previously shown, when sputtering is performed using the metallic Ni target, the AgNW networks remain structurally and electrically intact even in the absence of a buffer layer. Grazing-incidence XRD patterns retain all characteristic Ag reflections after deposition, while SEM observations reveal continuous, non-fragmented nanowires. Consistently, the electrical resistance of the network remains stable (Fig.~\ref{fig:coatings}(a,b)). For the sake of completeness, it should be noted that ZnO-buffered samples subjected to NiO sputtering also exhibit excellent structural and electrical stability, although this outcome is expected given the already benign nature of the NiO sputtering environment.
In sharp contrast and similarly to the results obtained with the W target, deposition using the WO$_3$ target without any buffer layer results in the degradation of the AgNW network, even in the absence of added oxygen (Fig.~\ref{fig:coatings}c). The Ag diffraction peaks disappear from the XRD pattern, SEM images show severe nanowire fragmentation, and the electrical resistance diverges to infinity. This behavior is consistent with the efficient generation of energetic oxygen-related negative ions from the oxidized tungsten surface, which are known to induce strong resputtering effects at the substrate.

Remarkably, inserting a thin ZnO buffer layer (30~nm) prior to WO$_3$ deposition fully suppresses this degradation (Fig.~\ref{fig:coatings}(a,e)). Under otherwise identical sputtering conditions, the Ag diffraction peaks remain clearly visible, SEM imaging shows continuous nanowires without evidence of fragmentation, and the network resistance remains finite. The ZnO interlayer therefore acts as an effective barrier against ion-induced damage, most likely by absorbing or scattering energetic species before they reach the fragile AgNW percolation network.

These results highlight the critical and previously underexplored role of ultrathin oxide buffer layers in stabilizing fragile percolating metal nanowire networks during reactive sputter deposition. This buffer-layer strategy provides a simple and scalable route to integrate AgNW electrodes with sputtered functional oxides while preserving their electrical performance.

The key findings of this study are summarized in Table~1. They clearly show that the presence of oxygen is a necessary but not sufficient condition for AgNW degradation during RF magnetron sputtering. The chemical nature of the sputtering target plays a decisive role, with the degradation severity qualitatively following the effective oxygen affinity of the target material. Targets that readily form stable surface oxides, such as W and WO$_3$, systematically induce severe network failure, whereas metallic targets with lower effective oxidation tendencies, such as Cu and Ni, produce little to no degradation under comparable conditions. These observations indicate that the efficiency of degradation is governed by the coupled oxygen–target interaction rather than by oxygen concentration alone, consistent with a mechanism involving energetic negative oxygen ions generated at oxidized target surfaces. Notably, the strong degradation observed for metallic W under reactive conditions further supports the role of rapid target poisoning, whereby the surface effectively behaves as an oxide during sputtering. Importantly, in all tested configurations, the introduction of a 30~nm ZnO buffer layer effectively suppresses degradation, which is consistent with a predominantly ballistic ion-induced damage mechanism in which the buffer layer acts as a physical shield against energetic oxygen species.

\section{Conclusions}
The integration of AgNW networks into multilayer functional devices requires deposition processes that preserve both their structural integrity and electrical performance. Although RF magnetron sputtering is one of the most versatile and scalable thin film deposition techniques, its interaction with fragile percolative nanowire networks has remained insufficiently clarified. Establishing robust processing guidelines is therefore critical for the reliable implementation of AgNW electrodes in optoelectronic, photovoltaic, and sensing platforms.

In this work, we investigated the degradation of AgNW networks during RF magnetron sputtering over a broad range of conditions, with particular emphasis on oxygen partial pressure and target chemistry. Our results demonstrate that the observed damage is driven by species generated at the target when oxygen is present in the discharge. In oxygen-free atmospheres, AgNW networks remain structurally and electrically stable across the examined conditions, identifying oxygen as a key enabling factor for degradation.

When oxygen is introduced, the nature of the sputtering target becomes decisive. Oxide targets, as well as metallic targets that readily oxidize under reactive conditions, systematically induce severe structural fragmentation and electrical failure of the AgNW networks. The trends are consistent with a damage mechanism dominated by energetic negative oxygen ions originating from the target surface. In contrast, metallic targets with lower effective oxidation propensity, such as Ni and Cu, enable stable coatings within a limited processing window. Importantly, mitigation strategies commonly effective for compact thin films—such as increasing the argon pressure, reducing deposition time, or lowering plasma power—prove insufficient for nanowire networks, highlighting the intrinsic vulnerability of percolative nanostructures to ion-induced damage.

The introduction of thin oxide buffer layers prior to sputtering emerges as a robust and versatile strategy to preserve AgNW integrity. These interlayers efficiently suppress network fragmentation even under otherwise damaging reactive sputtering conditions, providing a practical pathway toward the integration of AgNW electrodes with sputtered functional oxides.

In conclusion, this work establishes a mechanistic framework linking oxygen presence, target reactivity, and AgNW degradation during RF magnetron sputtering. More broadly, it demonstrates that processing rules developed for continuous thin films cannot be directly transposed to nanostructured networks. Future efforts should focus on quantitatively correlating negative ion fluxes with target properties and plasma conditions, as well as on developing plasma engineering strategies specifically tailored to nanowire-based electrodes. Such advances will be essential to enable the reliable, large-scale integration of AgNW networks into next-generation multilayer devices.

\begin{acknowledgments}
The authors gratefully acknowledge the financial support from the European Commission via the M-ERA.NET program (INSTEAD project). A. B. and N. D. N. acknowledge the financial support from F.R.S. – FNRS via the PINT-MULTI project R.8012.20 and the CDR project J.0157.24.
\end{acknowledgments}
\bibliography{Plasmafinal}
\newpage
\pagebreak
\begin{center}
\textbf{\large Supplementary Materials}
\end{center}
\setcounter{equation}{0}
\setcounter{figure}{0}
\setcounter{table}{0}
\setcounter{page}{1}
\setcounter{section}{0}

\makeatletter
\renewcommand{\theequation}{S\arabic{equation}}
\renewcommand{\thefigure}{S\arabic{figure}}
\renewcommand{\bibnumfmt}[1]{[S#1]}
\renewcommand{\citenumfont}[1]{S#1}

\section{On the influence of argon pressure}

While the generation of etch-inducing negative oxygen ions originates from interactions between oxygen and the target material—and can therefore, in principle, be mitigated by modifying either the oxygen content or the target surface state—an alternative strategy consists in increasing the total argon pressure. Raising the inert gas pressure increases the probability of gas-phase collisions involving negative oxygen ions, thereby reducing their mean free path and kinetic energy upon reaching the substrate.

\begin{figure*}
    \centering
    \includegraphics[width=0.85\linewidth]{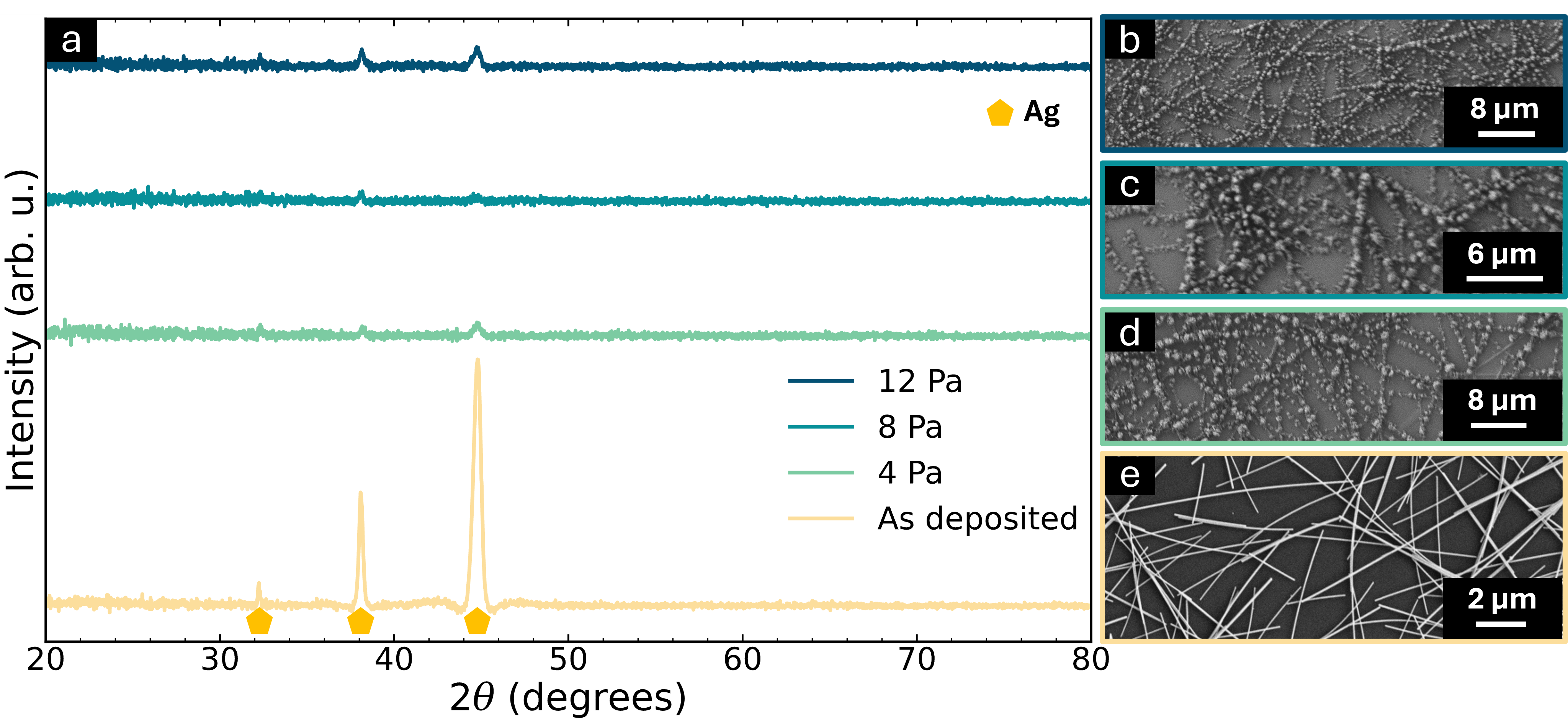}
    \caption{\textit{(a)} Grazing incidence XRD patterns of AgNW networks on glass before and after RF magnetron sputtering performed at a plasma power of 20~W and an oxygen partial pressure of 12\%. 
\textit{(b-d)} SEM images of AgNW networks after sputtering at total argon pressures of 12~Pa, 8~Pa, and 4~Pa, respectively. 
\textit{(e)} SEM image of a bare AgNW network prior to sputtering, shown for reference.
}
    \label{fig:ArPressureXRDSEM}
\end{figure*}

The corresponding results are presented in Fig.~\ref{fig:ArPressureXRDSEM}. The experiments were performed at an RF power of 20~W with 12\% O$_2$ partial pressure, using a metallic W target. Although tungsten is metallic, it is characterized by a high oxide formation enthalpy, indicating a strong affinity for oxygen. As a result, tungsten readily forms a thin surface oxide under oxygen-containing discharges (see Table~\ref{tab:summary_targets}), which may promote the generation of negative oxygen ions despite the metallic nature of the target. The plasma power was kept at the minimum value required to sustain a stable discharge and to obtain WO$_3$ films, while the argon pressure was varied from 4 to 12 Pa.

The results show that increasing the argon pressure to 12~Pa results in only a marginal increase in the Ag diffraction peak intensity compared to lower-pressure conditions, indicating that structural degradation of the AgNW network is not effectively mitigated. Consistently, the electrical resistance (infinite after sputtering) further confirms that increasing the argon pressure alone is insufficient to suppress degradation under the investigated sputtering conditions.

Moreover, such elevated argon pressures are inherently associated with reduced deposition rates and, consequently, longer deposition times. When combined with the absence of a clear protective effect, this renders pressure-based mitigation impractical for preserving AgNW integrity. It can therefore be concluded that increasing the argon pressure and decreasing the plasma power are not, by themselves, an effective strategy to prevent AgNW degradation during RF magnetron sputtering, and that alternative mitigation pathways must be explored.

Nevertheless, this observation remains significant, as pressure-based shielding is a commonly employed strategy in oxide sputtering processes. Its limited effectiveness in the present context highlights the need for a more nuanced understanding of ion-induced degradation mechanisms when coating fragile nanostructured substrates.

\end{document}